\documentclass[%
reprint,
superscriptaddress,
showpacs,
amsmath,amssymb,
prc,
floatfix ]%
{revtex4-1}

\usepackage{color}
\usepackage{CJK}

\usepackage{graphicx}% Include figure files
\usepackage{dcolumn}% Align table columns on decimal point
\usepackage{bm}% bold math
\usepackage{hyperref}% add hypertext capabilities
\allowdisplaybreaks

\begin{document}

\begin{CJK*}{UTF8}{}
\title{Multidimensionally-constrained relativistic mean-field study 
of spontaneous fission: coupling between shape and pairing degrees of freedom}
\CJKfamily{gbsn}
\author{Jie Zhao (赵杰)}%
%\author{Jie Zhao}
\email{zhaojie@itp.ac.cn}
 \affiliation{Physics Department, Faculty of Science, University of Zagreb, Bijeni\v{c}ka Cesta 32,
              Zagreb 10000, Croatia}
% \affiliation{State Key Laboratory of Theoretical Physics,
%              Institute of Theoretical Physics, Chinese Academy of Sciences, Beijing 100190, China}
\CJKfamily{gbsn}
\author{Bing-Nan Lu (吕炳楠)}%
%\author{Bing-Nan Lu}
%\email{zhaojie@itp.ac.cn}
 \affiliation{Institute for Advanced Simulation, Institut f\"ur Kernphysik,
	      and J\"ulich Center for Hadron Physics,
              Forschungszentrum J\"ulich, D-52425 J\"ulich, Germany}
% \affiliation{State Key Laboratory of Theoretical Physics,
%              Institute of Theoretical Physics, Chinese Academy of Sciences, Beijing 100190, China}
\author{Tamara Nik\v{s}i\'c}%
%\email{tniksic@phy.hr}
 \affiliation{Physics Department, Faculty of Science, University of Zagreb, Bijeni\v{c}ka Cesta 32,
              Zagreb 10000, Croatia}             
\author{Dario Vretenar}%
%\email{vretenar@phy.hr}
 \affiliation{Physics Department, Faculty of Science, University of Zagreb, Bijeni\v{c}ka Cesta 32,
              Zagreb 10000, Croatia}
\CJKfamily{gbsn}              
\author{Shan-Gui Zhou (周善贵)}%
%\email{sgzhou@itp.ac.cn}
 \affiliation{State Key Laboratory of Theoretical Physics, Institute of Theoretical Physics, Chinese Academy of Sciences, Beijing 100190, China}
 \affiliation{Center of Theoretical Nuclear Physics, National Laboratory of Heavy Ion Accelerator, Lanzhou 730000, China}
 \affiliation{Synergetic Innovation Center for Quantum Effects and Application, Hunan Normal University, Changsha 410081, China}

\date{\today}

\begin{abstract}
\begin{description}
\item[Background]
Studies of fission dynamics, based on nuclear energy density functionals, have shown that the coupling between 
shape and pairing degrees of freedom has a pronounced effect on the nonperturbative collective 
inertia and, therefore, on dynamic (least-action) spontaneous fission paths and half-lives.  
\item[Purpose] To analyze effects of particle-number fluctuation degree of freedom on symmetric and asymmetric 
spontaneous fission (SF) dynamics, and compare with results of recent studies based 
on the self-consistent Hartree-Fock-Bogoliubov (HFB) method.
\item[Methods]
Collective potentials and nonperturbative cranking collective inertia tensors 
are calculated using the multidimensionally-constrained relativistic mean-field (MDC-RMF) model.
Pairing correlations are treated in the BCS approximation using a separable pairing force of finite range. 
Pairing fluctuations are included as a collective variable using a constraint on particle-number dispersion.
Fission paths are determined with the dynamic programming method 
by minimizing the action in multidimensional collective spaces.
\item[Results]
The dynamics of spontaneous fission of $^{264}$Fm and $^{250}$Fm are explored. 
Fission paths, action integrals and corresponding half-lives computed in the three-dimensional 
collective space of shape and pairing coordinates, using the relativistic functional DD-PC1 
and a separable pairing force of finite range, are compared with results obtained without pairing fluctuations.  
Results for $^{264}$Fm are also discussed in relation with those recently obtained using the HFB model. 
\item[Conclusions]
The inclusion of pairing correlations in the space of collective coordinates favors 
axially symmetric shapes along the dynamic path of the fissioning system, 
amplifies pairing as the path 
traverses the fission barriers, significantly reduces the action integral and shortens
the corresponding SF half-life. 
\end{description}
\end{abstract}

%21.60.Jz       Nuclear Density Functional Theory and extensions
%               (includes Hartree-Fock and random-phase approximations)
%24.75.+i       General properties of fission
%25.85.Ca 	Spontaneous fission
%27.90.+b       A >= 220

\maketitle

\end{CJK*}

\bigskip

\section{Introduction~\label{sec:Introduction}}

Important advances have recently been reported in microscopic modeling of the dynamics of spontaneous and 
induced fission, based on nuclear density functional theory (DFT) \cite{Schunck2015_arXiv}. Within this framework, 
spontaneous fission (SF), in particular, is described by quantum tunneling through potential barrier(s) in a multidimensional 
space of coordinates that parametrize large-amplitude collective motion. Most calculations of SF lifetimes are based on 
the semiclassical Wentzel-Kramers-Brillouin (WKB) approximation for the one-dimensional barrier tunneling. The dynamics 
of the SF process is governed by the potential energy surface (PES) as function of the collective coordinates, and by the 
collective inertia along the fission path. The path along which the nucleus tunnels is determined
by minimizing the fission action integral in the multidimensional collective space~\cite{Brack1972_RMP44-320,Ledergerber1973_NPA207-1}.

The PES's can be computed using the macroscopic-microscopic (MM) model,  
or a number of self-consistent mean-field (SCMF) approaches based on microscopic effective interaction or energy density functionals.
In most recent studies the multidimensional collective inertia tensor is usually determined using the
adiabatic time-dependent Hartree-Fock-Bogoliubov (ATDHFB) method with the 
perturbative cranking approximation (neglecting the contribution from time-odd mean fields and treats perturbatively the derivatives 
of single-nucleon and pairing densities with respect to collective coordinates), or the 
nonperturbative cranking approximation (the derivatives with respect to collective coordinates are computed explicitly).
Although perturbative cranking ATDHFB collective masses has extensively been used in SF fission half-life 
calculations~\cite{Brack1972_RMP44-320,Nilsson1969_NPA131-1,Girod1979_NPA330-40,Bes1961_NP28-42,
Sobiczewski1969_NPA131-67}, a number of recent studies \cite{Baran2011_PRC84-054321,Sadhukhan2013_PRC88-064314,Zhao2015_PRC92-064315} have indicated the 
essential role of the nonperturbative cranking ATDHFB approximation to the collective inertia for a quantitative 
dynamic description of SF.

In the first approximation the effective collective inertia $\mathcal{M} \propto \Delta^{-2}$, 
and the collective potential $V \propto (\Delta-\Delta_0)^2$, where $\Delta$ is the pairing gap and $\Delta_0$ 
corresponds to its self-consistent stationary value. When the gap parameter is treated as a dynamical variable, 
an enhancement of pairing correlations reduces the effective inertia and thus minimizes the  
action integral $S$ along the fission path \cite{Moretto1974_PLB49-147}. A number of studies 
of SF have shown that the coupling of pairing fluctuations with the fission mode can significantly reduce the estimated fission 
lifetimes~\cite{Urin1966_NP75-101,Lazarev1987_PS35-255,Staszczak1985_PLB161-227,Smolanczuk1993_APPB24-685,
Staszczak1989_NPA504-589,Lojewski1999_NPA657-134,Mirea2010_JPG37-055106,Pomorski2007_IJMPE16-237}. 

In recent studies dynamic fission paths determined with the least-action principle have been 
investigated using the Hartree-Fock-Bogogliubov (HFB) framework based on the 
Barcelona-Catania-Paris-Madrid~\cite{Giuliani2014_PRC90-054311}, Gogny D1M~\cite{Giuliani2014_PRC90-054311}, 
and Skyrme SkM$^*$~\cite{Sadhukhan2014_PRC90-061304,Sadhukhan2016_PRC93-011304} energy density functionals.
The pairing gap parameter has been included as a dynamical variable in the collective space. As a result, 
an enhancement of pairing correlations along fission paths and the speedup of SF have been predicted.
It has also been noted that pairing fluctuations can restore axial symmetry in the fissioning 
system \cite{Sadhukhan2014_PRC90-061304,Sadhukhan2016_PRC93-011304}, although 
the triaxial quadrupole degree of freedom is known to play an important role 
around the inner and even outer barriers both along the static fission path for actinide nuclei 
(Ref.~\cite{Lu2014_PRC89-014323} and references therein),  
and in the dynamic case when the influence of pairing fluctuations is not taken 
into account \cite{Sadhukhan2013_PRC88-064314,Zhao2015_PRC92-064315}. 

In Ref.~\cite{Zhao2015_PRC92-064315} we have used the 
multidimensionally-constrained relativistic Hartree-Bogoliubov 
(MDC-RHB) to analyse effects of triaxial and octupole deformations, as 
well as approximations to the collective inertia, on the symmetric and asymmetric 
spontaneous fission dynamics. Based on the framework of relativistic energy density functionals, and 
using as examples $^{264}$Fm and $^{250}$Fm, our analysis has 
shown that the action integrals and, consequently, the half-lives crucially depend 
on the approximation used to calculate the effective collective inertia along the fission path. 
While the perturbative cranking approach underestimates the effects of 
structural changes at the level crossings, the 
nonperturbative collective mass is characterized by the occurrence of sharp peaks 
on the surface of collective coordinates, that can be related to single-particle level crossings 
near the Fermi surface, and this enhances the effective inertia. 
In this work we continue to explore the dynamics of SF of $^{264}$Fm and $^{250}$Fm but, 
in addition to shape deformation degrees of freedom, pairing correlations are included in the  
space of collective coordinates. The dynamic (least-action) fission paths are determined in 
three-dimensional (3D) collective spaces, and the corresponding SF half-lives are computed.
Since calculations in the 3D collective space with the MDC-RHB model are computationally very demanding, 
here we employ the MDC-RMF model in which the pairing correlations are treated in the BCS approximation. 
The collective inertia tensor is calculated using the self-consistent relativistic mean-field (RMF) solutions and  
applying the ATDHFB expressions in the nonperturbative cranking approximation. 
The article is organized as follows: the method for calculating dynamic fission paths is 
described in Sec.~\ref{sec:model}; numerical details of the calculation, results for 
the deformation energy landscapes, collective inertias, minimum-action fission paths 
and the corresponding half-lives are discussed in Sec.~\ref{sec:results}; and 
Sec.~\ref{sec:summary} contains a short summary of the main results.

%%%%%%%%%%%%%%%%%%%%%%%%%%%%%%%%%%%%%%%%%%%%
\section{\label{sec:model}Method for calculating dynamic fission paths}
%%%%%%%%%%%%%%%%%%%%%%%%%%%%%%%%%%%%%%%%%%%%
RMF-based models present a particular implementation of the relativistic 
nuclear energy density functional (EDF) framework, which has become a standard 
method for studies of the structure of medium-heavy and heavy 
nuclei~\cite{Serot1986_ANP16-1,Reinhard1989_RPP52-439,Ring1996_PPNP37-193,
Bender2003_RMP75-121,Vretenar2005_PR409-101,Meng2006_PPNP57-470}.
As in our previous study of spontaneous fission \cite{Zhao2015_PRC92-064315}, 
here we employ the point-coupling relativistic EDF DD-PC1~\cite{Niksic2008_PRC78-034318}.  
Starting from microscopic nucleon self-energies in nuclear matter, 
and empirical global properties of the nuclear matter equation of state, the coupling parameters 
of DD-PC1 were fine-tuned to the experimental masses of a set of 64 deformed nuclei in the 
mass regions $A \approx 150 -180$ and $A \approx 230 - 250$. The functional has been further 
tested in a number of mean-field and beyond-mean-field calculations in different mass 
regions.

For a quantitative description of open-shell nuclei it is necessary to consider also pairing correlations. 
In the MDC-RMF model pairing is taken into account in the BCS approximation and here, 
as in Ref.~\cite{Zhao2015_PRC92-064315}, we use a separable
pairing force of finite range:
\begin{equation}
V(\mathbf{r}_1,\mathbf{r}_2,\mathbf{r}_1^\prime,\mathbf{r}_2^\prime) = G_0 ~\delta(\mathbf{R}-
\mathbf{R}^\prime) P (\mathbf{r}) P(\mathbf{r}^\prime) \frac{1}{2} \left(1-P^\sigma\right),
\label{pairing}
\end{equation}
where $\mathbf{R} = (\mathbf{r}_1+\mathbf{r}_2)/2$ and $\mathbf{r}=\mathbf{r}_1- \mathbf{r}_2$
denote the center-of-mass and the relative coordinates, respectively, and $P(\mathbf{r})$ reads 
\begin{equation}
P(\mathbf{r})=\frac{1}{\left(4\pi a^2\right)^{3/2}} e^{-\mathbf{r}^2/4a^2}.
\end{equation}
The two parameters $G_0=-738$ MeV fm$^{-3}$ and $a=0.644$ fm~\cite{Tian2009_PLB676-44}
have been adjusted to reproduce the density dependence of the pairing gap in nuclear matter at the
Fermi surface calculated with the D1S parameterization of the Gogny force~\cite{Berger1991_CPC63-365}.

The energy landscape is obtained in a self-consistent mean-field 
calculation with constraints on mass multipole moments $Q_{\lambda\mu} = r^\lambda Y_{\lambda \mu}$, 
and the particle-number dispersion operator $\Delta\hat{N}^2 = \hat{N}^2 - \langle \hat{N} \rangle^2$~\cite{Ring1980}. 
In the present analysis the Routhian is therefore defined as
\begin{equation}
E^\prime = E_{\rm RMF} + \sum_{\lambda\mu}{\frac{1}{2}C_{\lambda \mu}Q_{\lambda \mu}} + \lambda_{2} \Delta\hat{N}^2 \;,
\end{equation}
where $E_{\rm RMF}$ denotes the total RMF energy including static BCS pairing correlations. 
The amount of dynamic pairing correlations can be controlled by the Lagrange multipliers 
$\lambda_{2\tau}$ ($\tau=n,p$),~\cite{Vaquero2011_PLB704-520,Vaquero2013_PRC88-064311,Sadhukhan2014_PRC90-061304}.
As it has recently been shown in a similar study of Ref.~\cite{Sadhukhan2014_PRC90-061304}, 
the isovector pairing degree of freedom appears to play a far less important role in spontaneous fission 
as compared to isoscalar dynamic pairing. Therefore, the computational task can be greatly reduced by considering only 
dynamic pairing with $\lambda_{2n} = \lambda_{2p} \equiv \lambda_{2}$ as a collective coordinate.
%---------------------------

The nuclear shape is parameterized by the deformation parameters
\begin{equation}
 \beta_{\lambda\mu} = {4\pi \over 3AR^\lambda} \langle Q_{\lambda\mu} \rangle.
\end{equation}
The shape is assumed to be invariant under the exchange of the $x$ and $y$ axes 
and all deformations $\beta_{\lambda\mu}$ with even $\mu$ can be included simultaneously.
The deformed RMF equations are solved by an expansion in the 
axially deformed harmonic oscillator (ADHO) basis~\cite{Gambhir1990_APNY198-132}.
In the present study of transactinide nuclei calculations have been performed 
in an ADHO basis truncated to $N_f = 16$ oscillator shells.
For details of the MDC-RMF model we refer the reader to Ref.~\cite{Lu2014_PRC89-014323}.

The action integral along the one-dimensional fission path $L$ is calculated using the expression:
\begin{equation}
\label{eq:act_integration}
S(L) = \int_{s_{\rm in}}^{s_{\rm out}} {1\over\hbar} 
  \sqrt{ 2\mathcal{M}_{\rm eff}(s) \left[ V_{\rm eff}(s)-E_0 \right] } ds \;,
\end{equation}
where $\mathcal{M}_{\rm eff}(s)$ and $V_{\rm eff}(s)$ are the effective 
collective inertia and potential along the path $L(s)$, respectively.
$E_0$ is the collective ground state energy, and the integration
limits correspond to the classical inner ($s_{\rm in}$) and outer ($s_{\rm out}$) turning points defined by: 
$V_{\rm eff}(s) = E_0$.
The fission path $L(s)$ is determined in the multidimensional collective space 
by minimizing the action integral of Eq.~(\ref{eq:act_integration})~\cite{Brack1972_RMP44-320,Ledergerber1973_NPA207-1}.
The spontaneous fission half-life is calculated as $T_{1/2}=\ln2/(nP)$, where $n$ is the
number of assaults on the fission barrier per unit 
time~\cite{Baran1978_PLB76-8,Baran1981_NPA361-83,
Sadhukhan2013_PRC88-064314,Sadhukhan2014_PRC90-061304}, 
and $P$ is the barrier penetration probability in the WKB approximation
\begin{equation}
P = {1 \over 1+\exp[2S(L)]}. 
\end{equation}

The action integral Eq. (\ref{eq:act_integration}) and, therefore, the fission half-life is essentially determined by the 
effective collective inertia and potential. The effective inertia is defined in terms of the 
multidimensional collective inertia tensor $\mathcal{M}$
\cite{Brack1972_RMP44-320,Baran1978_PLB76-8,Baran1981_NPA361-83,Sadhukhan2013_PRC88-064314,
Sadhukhan2014_PRC90-061304,Zhao2015_PRC92-064315}
\begin{equation}
\mathcal{M}_{\rm eff}(s) = \sum_{ij} \mathcal{M}_{ij} {dq_i \over ds} {dq_j \over ds}\;,
\label{inertia}
\end{equation}
where $q_i(s)$ denotes the collective variable as function of the path's length. 

In the present study the inertia tensor is computed using the ATDHFB method in 
the nonperturbative cranking approximation~\cite{Baran2011_PRC84-054321} 
\begin{equation}
\label{eq:npmass}
\mathcal{M}_{ij}^{C} = {\hbar^2 \over 2 \dot{q}_i \dot{q}_j}
    \sum_{\alpha\beta} {F^{i*}_{\alpha\beta}F^{j}_{\alpha\beta} + F^{i}_{\alpha\beta}F^{j*}_{\alpha\beta}
    \over E_{\alpha} + E_{\beta}},
\end{equation}
where
\begin{equation}
\label{eq:fmatrix}
{F^{i} \over \dot{q}_{i}}  
  =  U^\dagger {\partial\rho \over \partial q_{i}} V^* 
    + U^\dagger {\partial\kappa \over \partial q_{i}} U^*
    - V^\dagger {\partial\rho^* \over \partial q_{i}} U^*
    - V^\dagger {\partial\kappa^* \over \partial q_{i}} V^*\;.
\end{equation}
$U$ and $V$ are the self-consistent Bogoliubov matrices, and $\rho$ and $\kappa$ are 
the corresponding particle and pairing density matrices, respectively.
The derivatives of the densities are calculated using the Lagrange three-point formula for 
unequally spaced points~\cite{Giannoni1980_PRC21-2076,Yuldashbaeva1999_PLB461-1}.

The collective potential $V_{\rm eff}$ is obtained by subtracting the vibrational zero-point energy (ZPE) from the total RMF
constrained energy surface \cite{Staszczak2013_PRC87-024320,Baran2007_IJMPE16-443,Sadhukhan2013_PRC88-064314,
Sadhukhan2014_PRC90-061304,Zhao2015_PRC92-064315}. 
The fission path is determined in a multidimensional collective space using both the dynamic programming (DPM)~\cite{Baran1981_NPA361-83}
and Ritz ~\cite{Baran1978_PLB76-8} (RM) methods .
For both methods we have considered several possible values
for the turning points $s_{\rm in}$ and $s_{\rm out}$ to verify that
the minimum action path is chosen. Since both methods give virtually identical results, only those obtained using the DPM 
are included in the presentation.

%----------------------------------------------------------------------------------------------------------------------
\section{\label{sec:results}Spontaneous fission of $^{\bf 264}{\rm \bf Fm}$ and $^{\bf 250}{\rm \bf Fm}$: 
pairing-induced speedup}
%----------------------------------------------------------------------------------------------------------------------
To study the effect of dynamic pairing correlations along fission paths, 
as in our previous study of SF of Ref.~\cite{Zhao2015_PRC92-064315},
we will analyze two illustrative examples: the symmetric spontaneous fission of 
$^{264}$Fm and the asymmetric SF of $^{250}$Fm. In addition to shape variables, 
here pairing correlations are also considered as collective coordinates in the study of 
fission dynamics. Because of computational restrictions and to simplify 
the interpretation of results, the present analysis is restricted to a 
three-dimensional (3D) collective space, defined by either
$(\beta_{20},\beta_{22},\lambda_{2})$ (quadrupole triaxial shapes) 
or $(\beta_{20},\beta_{30},\lambda_{2})$ (quadrupole and octupole axial shapes), 
where the coordinate $\lambda_{2}$ represents dynamic pairing fluctuations.
The relativistic energy density functional DD-PC1 \cite{Niksic2008_PRC78-034318}
is employed in self-consistent RMF calculations of constrained energy surfaces, 
collective inertia tensors and fission action integrals. The height of fission barriers is sensitive to
the strength of the pairing interaction~\cite{Karatzikos2010_PLB689-72} and, therefore, 
a particular choice of the pairing strength may have a considerable effect on fission dynamics.
As explained above and in Ref.~\cite{Zhao2015_PRC92-064315}, the parameters of 
the finite range separable pairing force were originally adjusted to reproduce the pairing gap at the 
Fermi surface in symmetric nuclear matter as calculated with the Gogny force D1S.
A number of mean-field studies based on the relativistic Hartree-Bogoliubov (RHB) model
have shown that the pairing strength needs to be fine-tuned in some 
cases, especially for heavy nuclei~\cite{Wang2013_PRC87-054331,Afanasjev2013_PRC88-014320}.
Since in the present study pairing correlations are treated in the BCS approximation, we have 
adjusted the strength parameters to reproduce the available empirical pairing gaps in Fm isotopes.
The resulting values with respect to the original 
pairing strength adjusted in nuclear matter ($G_0=-738$ MeV fm$^{-3}$) are
$G_n/G_0=1.21$, $G_p/G_0=1.14$.
As in Refs.~\cite{Sadhukhan2013_PRC88-064314,Sadhukhan2014_PRC90-061304} 
and our previous work Ref.~\cite{Zhao2015_PRC92-064315},
we choose $E_0 =1$ MeV in Eq.~(\ref{eq:act_integration}) for the value of the collective ground state energy. 
This value enables a direct comparison of our results with those reported in previous studies, especially 
in Ref.~\cite{Sadhukhan2014_PRC90-061304}. 
For the vibrational frequency $\hbar\omega_0=1$ MeV the number of assaults on the fission barrier per unit 
is $10^{20.38}$ s$^{-1}$~\cite{Baran2005_PRC72-044310}.

%-----------------------------------------
\subsection{\label{subsec:Fm264}Symmetric fission of $^{\bf 264}{\rm \bf Fm}$}
%-----------------------------------------
%----
\begin{figure}
 \includegraphics[width=0.48\textwidth]{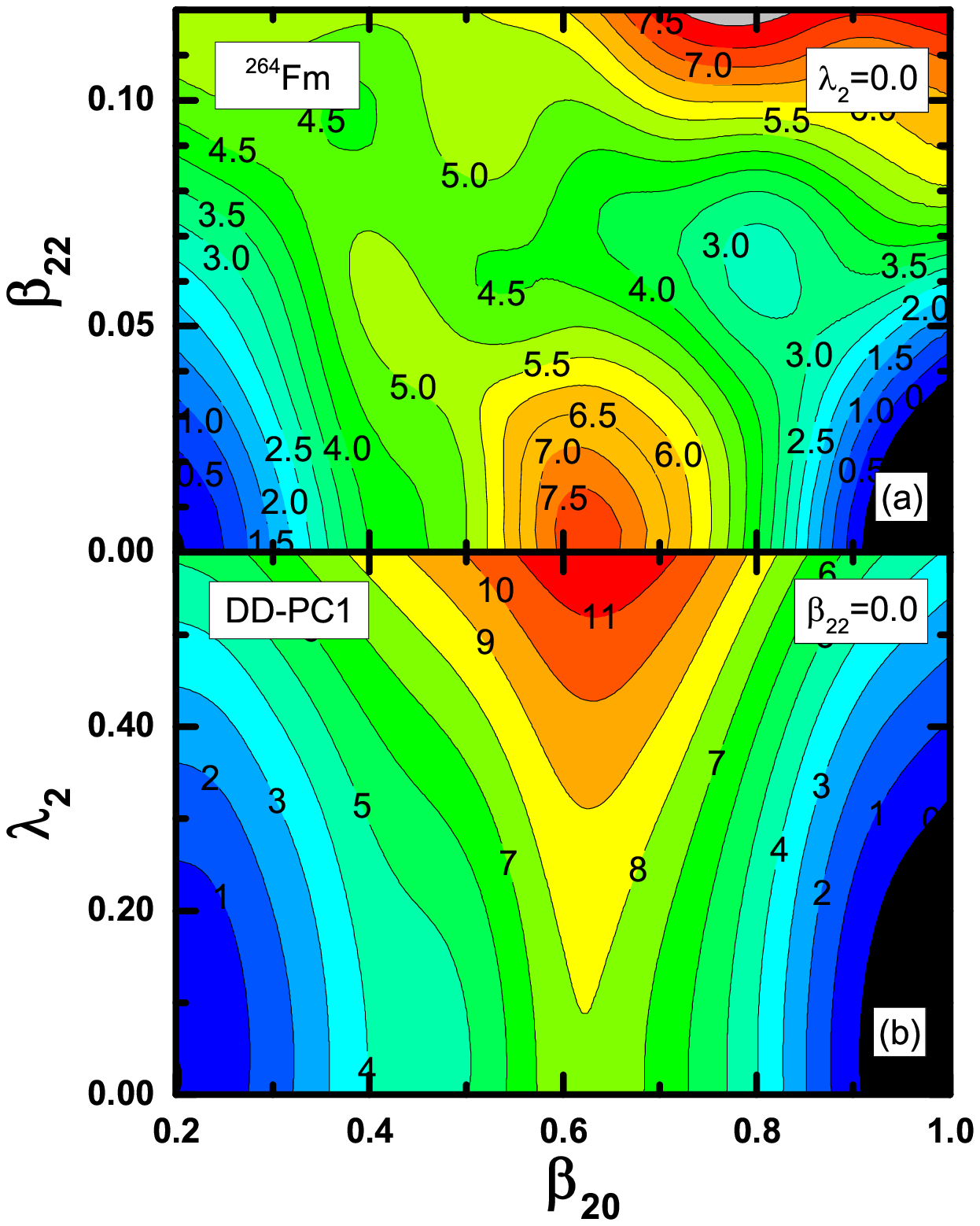}
\caption{(Color online)~\label{fig:Fm264_V}%
Effective collective potential $V_{\rm eff}$ of $^{264}$Fm in the $(\beta_{20},\beta_{22})$ plane for $\lambda_2 = 0$ (a),  
and in the $(\beta_{20},\lambda_2)$ plane for $\beta_{22}=0$ (b).
In each panel energies are normalized with respect to the corresponding value at the equilibrium minimum, 
and contours join points on the surface with the same energy (in MeV). 
The energy surfaces are calculated with the relativistic density functionals DD-PC1~\cite{Niksic2008_PRC78-034318}, 
and the pairing interaction Eq.~(\ref{pairing}).
}
\end{figure}
%-----
The first example in our analysis of the influence of dynamical fluctuations in shape and pairing degrees of freedom 
on fission paths is the nucleus $^{264}$Fm, for which theoretical studies 
\cite{Staszczak2009_PRC80-014309,Staszczak2011_IJMPE20-552} predict 
a symmetric spontaneous fission decay. The shape degrees of freedom in this case are 
elongation and triaxiality and, therefore, calculations of the energy landscape, inertia 
tensor, and fission paths are restricted to the 3D collective space $(\beta_{20},\beta_{22},\lambda_{2})$. 

In Fig. \ref{fig:Fm264_V} we plot the collective potential energy (the vibrational ZPE is subtracted from the 
constrained self-consistent mean-field energy) 
of $^{264}$Fm in the $(\beta_{20},\beta_{22})$ plane for $\lambda_{2} = 0$ (a),  
and in the $(\beta_{20},\lambda_{2})$ plane for $\beta_{22}=0$ (b). Therefore, panel (a) 
displays the results obtained without including dynamical pairing correlations, and the 
potential can be directly compared to the one shown in Fig. 6 (b) of Ref.~\cite{Zhao2015_PRC92-064315}, 
where we used the relativistic Hartree-Bogoliubov (RHB) model to compute the energy surface. 
We notice that the energy landscapes obtained with RMF and RHB models are almost identical, and this 
validates the treatment of pairing in the BCS approximation in the present analysis. The functional 
DD-PC1 predicts an axially symmetric 
equilibrium state at moderate deformation ($\beta_{20} \approx 0.2$), and 
the axially symmetric barrier at $\beta_{20}\approx 0.6$ is bypassed
through the triaxial region, thus lowering the height of the barrier by $\approx 2.5 - 3$ MeV.
In panel (b) of Fig.~\ref{fig:Fm264_V} we project the potential energy calculated in the 3D collective space 
that includes dynamic pairing, on the $(\beta_{20},\lambda_{2})$ plane. For $\beta_{22}=0$ (axially 
symmetric shape), the potential energy increases monotonically with $\lambda_{2}$ (stronger pairing) 
at each deformation $\beta_{20}$. The topography of the collective potential in the $(\beta_{20},\lambda_{2})$ plane
is relatively simple.  

%----
\begin{figure}
 \includegraphics[width=0.48\textwidth]{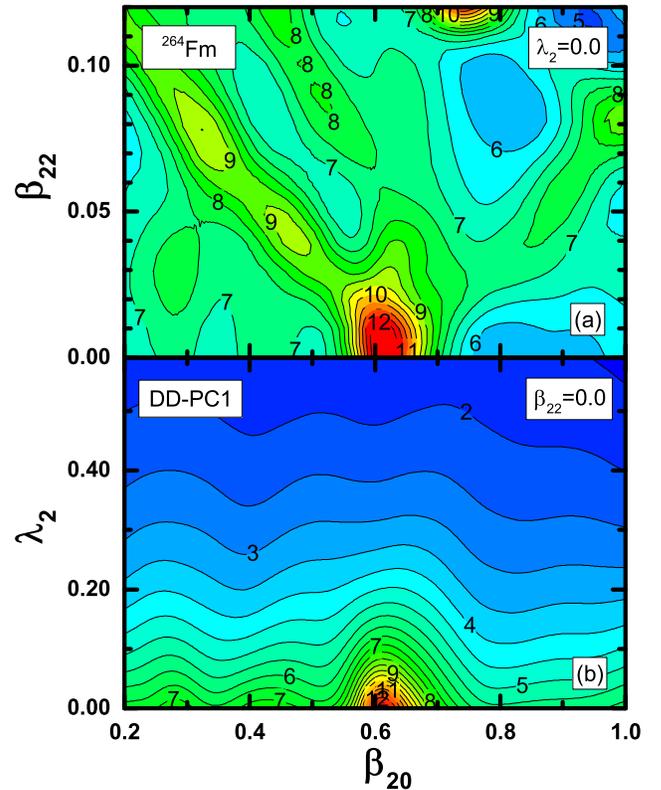}
\caption{(Color online)~\label{fig:Fm264_M}%
Cubic root determinants of the nonperturbative-cranking inertia tensor $|\mathcal{M}^{C}|^{1/3}$ 
(in $10 \times \hbar^2$MeV$^{-1}$) 
of $^{264}$Fm in the $(\beta_{20},\beta_{22})$ plane for $\lambda_{2} = 0$ (a),  
and in the $(\beta_{20},\lambda_2)$ plane for $\beta_{22}=0$ (b).
}
\end{figure}
%----

The effective inertia that determines the fission action integral is defined in terms of the 
multidimensional collective inertia tensor $\mathcal{M}$ Eq.~(\ref{inertia}).
The important effects related to the exact treatment of 
derivatives of single-particle and pairing densities in the ATDHFB expressions 
for the mass parameters were recently analyzed in Refs. \cite{Sadhukhan2013_PRC88-064314} 
and \cite{Zhao2015_PRC92-064315}. For the three-dimensional space of collective coordinates,  
six independent components determine the inertia tensor. 
The inertia tensor can be visualized by plotting the cubic root determinant $|\mathcal{M}|^{1/3}$,
invariant with respect to rotations in the three-dimensional collective space~\cite{Sadhukhan2014_PRC90-061304}.

In Fig.~\ref{fig:Fm264_M} we plot $|\mathcal{M}^{C}|^{1/3}$ obtained in the nonperturbative cranking approximation, 
in the $(\beta_{20},\beta_{22})$ plane for $\lambda_{2} = 0$ (a), and in the $(\beta_{20},\lambda_2)$ plane for $\beta_{22}=0$ (b). 
These results correspond to the self-consistent solutions for the potential energy surfaces shown in Fig.~\ref{fig:Fm264_V}.
The nonperturbative  $|\mathcal{M}^{C}|^{1/3}$, calculated without dynamic pairing correlations, displays a rather 
complex structure in the $(\beta_{20},\beta_{22})$ plane ($\lambda_{2}=0$), as shown in panel (a).
In particular, very large values of $|\mathcal{M}^C|^{1/3}$ are calculated in the region of the axial fission barrier.
As discussed in Refs.~\cite{Baran2011_PRC84-054321,Sadhukhan2013_PRC88-064314,Zhao2015_PRC92-064315},
this is related to single-particle level crossings near the Fermi surface. The abrupt changes of occupied single-particle 
configurations lead to strong variations in the derivatives of densities in Eq.~(\ref{eq:fmatrix})
and, consequently, sharp peaks develop. When dynamic pairing correlations are included in the 
collective space (panel (b)), a simple dependence of the nonperurbative 
$|\mathcal{M}^C|^{1/3}$ on $\lambda_2$ is obtained at each deformation $\beta_{20}$, 
consistent with the expected relation $\mathcal{M} \propto \Delta^{-2}$. We note that the results for 
$^{264}$Fm, shown in Figs.~\ref{fig:Fm264_V} and \ref{fig:Fm264_M}, are very similar to those obtained 
using the nonrelativistic HFB framework based on the Skyrme energy density functional SkM* and a 
density-dependent pairing interaction (cf. Fig. 2 of Ref.~\cite{Sadhukhan2014_PRC90-061304}). 

%----
\begin{figure}
 \includegraphics[width=0.48\textwidth]{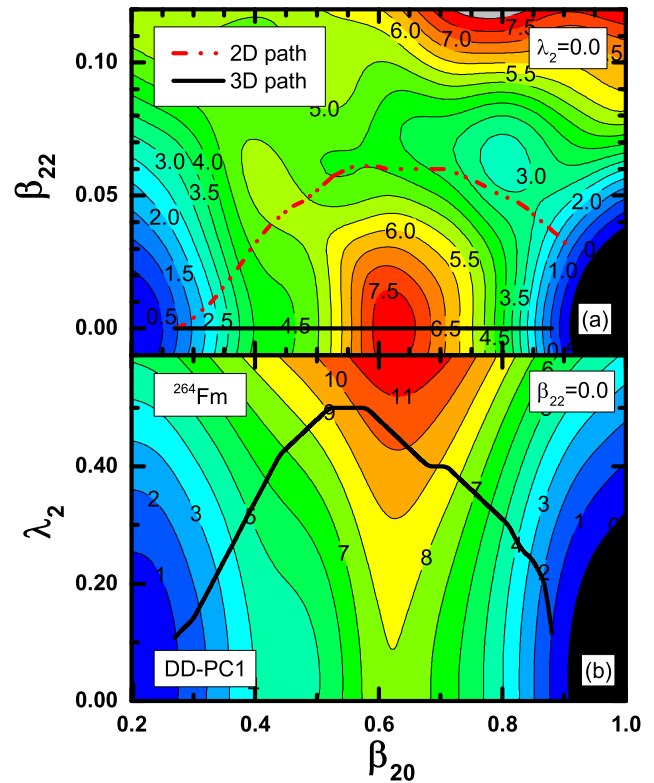}
\caption{(Color online)~\label{fig:Fm264_Path}%
Projections of the three-dimensional (3D) dynamic path (solid curves) for the spontaneous fission 
of $^{264}$Fm on the $(\beta_{20},\beta_{22})$ plane for $\lambda_2 = 0$ (a),  
and the $(\beta_{20},\lambda_2)$ plane for $\beta_{22}=0$ (b),
calculated using the dynamic programming method.
The dash-dot-dot curve denotes the two-dimensional (2D) path computed 
without the inclusion of dynamic pairing correlations. 
}
\end{figure}
%----
The coupling between shape and pairing degrees of freedom has a pronounced effect on the 
predicted fission paths. As the effective potential increases from the self-consistent values because of the 
enhancement of pairing, the effective collective inertia is reduced $\propto \Delta^{-2}$. 
These two effects determine the minimum-action path in Eq.~(\ref{eq:act_integration}). 
The projections of the 3D spontaneous fission path of $^{264}$Fm on the 
$(\beta_{20},\beta_{22})$ plane and on the $(\beta_{20},\lambda_{2})$ plane are shown 
in Fig.~\ref{fig:Fm264_Path} (a) and (b), respectively (solid curves). 
The two-dimensional (2D) path calculated without pairing fluctuations ($\lambda_{2}=0$)
is also included for comparison (dash-dot curve).
It is very interesting to note that, while  
the 2D dynamic path detours the axial barrier through the triaxial region, 
the extension of the collective space by the pairing degree of freedom fully 
restores the axial symmetry of the fissioning system. The evolution of the pairing 
strength along the axially symmetric fission path is shown in the lower panel of Fig.~\ref{fig:Fm264_Path}.
One notices how, in order to reduce the collective inertia, the fissioning nucleus 
favors an increase in pairing over the static self-consistent solution, at the expense of a 
larger potential energy. Because of pairing fluctuations, the 
corresponding fission action integral is reduced by about 5 units with respect to the 
2D path and, consequently, the predicted half-life is almost five orders of magnitude shorter in comparison to the 
2D case without the dynamic pairing degree of freedom (cf. Table~\ref{tab:act}). This result 
can directly be compared to the one obtained using the the Skyrme energy density functional SkM* and a 
density-dependent pairing interaction (cf. Fig. 3 of Ref.~\cite{Sadhukhan2014_PRC90-061304}). In the latter 
case triaxiality is reduced along the 3D fission path because of dynamic pairing fluctuations, but the full 
axial symmetry is not restored. 
%and, consequently, the calculated SF half-life is reduced by only 
%three orders of magnitude with respect to the 2D path. 
This is probably due to the fact that in the 
2D calculation with the Skyrme functional the triaxial coordinate reduces the fission barrier 
height by more than 4 MeV (less than 3 MeV in the present calculation with DD-PC1). A combination of 
a higher axially symmetric fission barrier and/or possibly weaker pairing, prevents the full restoration 
of axial symmetry along the 3D fission path of $^{264}$Fm. In the case of $^{240}$Pu, on the other hand, 
for which the Skyrme functional SkM* predicts an energy gain on the first barrier resulting from triaxiality of 
only 2 MeV, the inclusion of pairing fluctuations leads to a full restoration of axial symmetry along the 
3D fission path between the equilibrium ground state and the superdeformed fission isomer 
(cf. Fig. 5 of Ref.~\cite{Sadhukhan2014_PRC90-061304}).

%-----------------------------------------
\subsection{\label{subsec:Fm250}Asymmetric fission of $^{\bf 250}{\rm \bf Fm}$}
%-----------------------------------------

%----
\begin{figure}
 \includegraphics[width=0.48\textwidth]{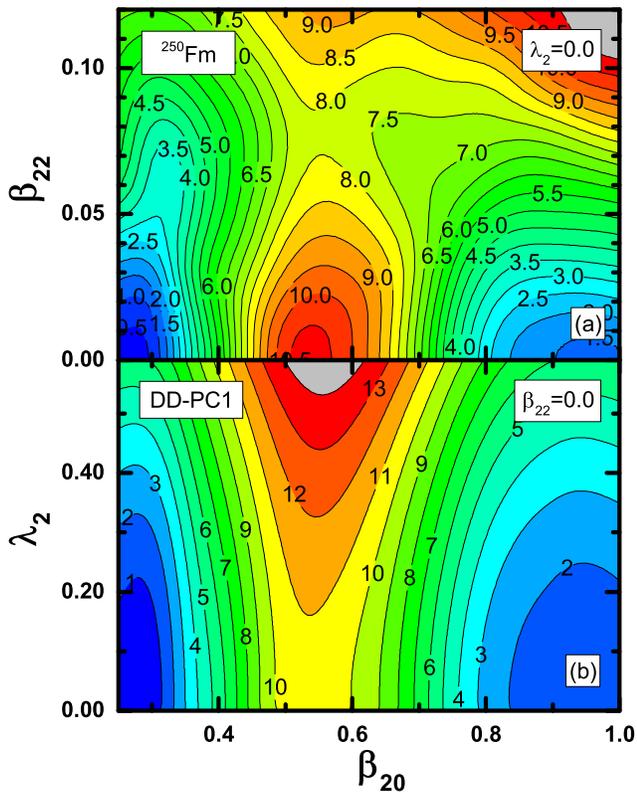}
\caption{(Color online)~\label{fig:Fm250_TFV}%
Same as described in the caption to Fig.~\ref{fig:Fm264_V} 
but for the nucleus $^{250}$Fm.
}
\end{figure}
%----

In the second example we explore the interplay between reflection-asymmetric shapes and pairing
degrees of freedom, and analyze the asymmetric spontaneous fission of $^{250}$Fm \cite{Staszczak2011_IJMPE20-552}.
Since the triaxial degree of freedom is particularly important around the inner fission barrier, 
and the complete calculation in the four-dimensional collective space 
($\beta_{20}$, $\beta_{22}$, $\beta_{30}$, $\lambda_{2}$) is computationally too demanding,
we first analyze the path that connects the mean-field equilibrium (ground) state and the isomeric fission 
state calculated in the $(\beta_{20},\beta_{22},\lambda_{2})$ collective space. 
The collective potential energy surface of $^{250}$Fm 
in the $(\beta_{20},\beta_{22})$  plane for $\lambda_{2}=0$, 
and in the $(\beta_{20},\lambda_{2})$ plane for $\beta_{22}=0$, 
is plotted in upper and lower panels of Fig.~\ref{fig:Fm250_TFV}, respectively.
The inclusion of the triaxial degree of freedom reduces the inner fission barrier height by $\approx$ 2 MeV, 
and this effect is similar in magnitude to the case of $^{264}$Fm considered in the previous section.
The lower panel displays the projection of the potential energy calculated in the 3D collective 
space on the $(\beta_{20},\lambda_{2})$ plane and we notice that, for $\beta_{22}=0$, the energy increases 
monotonically with $\lambda_{2}$ at each value of the axial deformation parameter $\beta_{20}$, with 
a pronounced fission barrier around $\beta_{20} \approx 0.55$.

%----
\begin{figure}
 \includegraphics[width=0.48\textwidth]{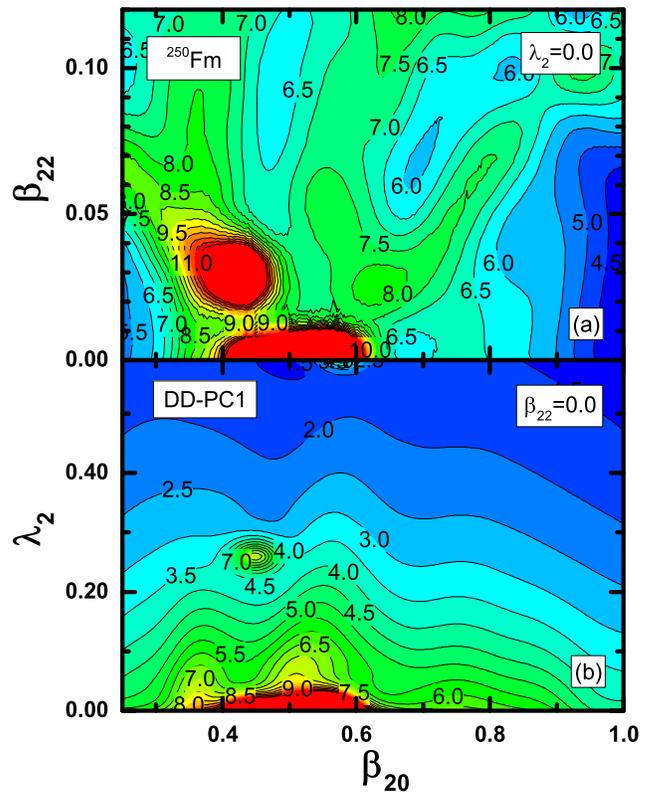}
\caption{(Color online)~\label{fig:Fm250_TFM}%
Same as described in the caption to Fig.~\ref{fig:Fm264_M} 
but for the nucleus $^{250}$Fm.
}
\end{figure}
%----

The deformation dependence of the nonperturbative collective inertia tensor is displayed in Fig.~\ref{fig:Fm250_TFM},
where we plot the cubic root determinants $|\mathcal{M}^{C}|^{1/3}$ in the $(\beta_{20},\beta_{22})$ 
and $(\beta_{20},\lambda_{2})$ planes. 
The global deformation dependence of $|\mathcal{M}^{C}|^{1/3}$ is similar to the one 
calculated for $^{264}$Fm and shown in Fig.~\ref{fig:Fm264_M}, that is, 
$|\mathcal{M}^{C}|^{1/3}$ displays strong variations in the ($\beta_{20},\beta_{22}$) plane for $\lambda_{2}=0$, 
and pronounced peaks generated by single-particle level crossings near the Fermi surface 
appear in the  region of the fission barrier. By including the dynamic pairing degree of freedom, 
one finds that $|\mathcal{M}^C|^{1/3}$ decreases as $\lambda_2$ 
increases at each deformation $\beta_{20}$.

%----
\begin{figure}
 \includegraphics[width=0.48\textwidth]{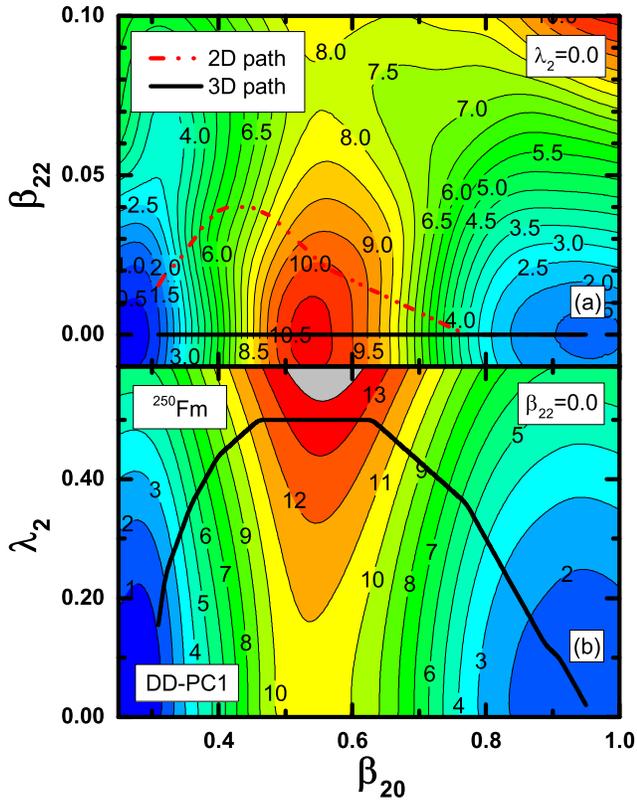}
\caption{(Color online)~\label{fig:Fm250_TF_Path}%
Projections of the three-dimensional (3D) dynamic fission path (solid curves)  
of $^{250}$Fm on the $(\beta_{20},\beta_{22})$ plane for $\lambda_2 = 0$ (a),  
and the $(\beta_{20},\lambda_2)$ plane for $\beta_{22}=0$ (b),
The dash-dot (red) curve denotes the two-dimensional (2D) path computed 
without the inclusion of dynamic pairing correlations. The minimum-action paths connect the 
inner turning point and the fission isomer.
}
\end{figure}
%----

The projections of the 3D dynamic path determined in the $(\beta_{20},\beta_{22},\lambda_{2})$ collective space
is shown in Fig.~\ref{fig:Fm250_TF_Path} (solid curves). The minimum-action path connects the  
inner turning point and the isomer minimum at $\beta_{20} \approx 0.95$. 
The 2D path calculated in $(\beta_{20},\beta_{22})$ collective space is also included (dash-dot red curve) for comparison. 
Even though the 2D dynamic path does not extend very far in the triaxial region, the 
triaxial shape degree of freedom is important in the calculation of the fission action integral, similar to the 
result we obtained with the RHB model in our previous study of fission dynamics (cf. Fig. 12 of \cite{Zhao2015_PRC92-064315}).
However, since triaxiality gains only $\approx$ 2 MeV in energy on the first barrier, the inclusion of 
dynamic pairing fluctuations fully restores axial symmetry in the fissioning system (upper panel of 
Fig.~\ref{fig:Fm250_TF_Path}). Pairing is enhanced with respect to the static solution along the 
axially symmetric path and, consequently, the action integral in the interval between the inner turning point and the isomeric 
state decreases from 23.06 (2D dynamic path) to 14.90 along the 3D fission path.

%----
\begin{figure}
 \includegraphics[width=0.48\textwidth]{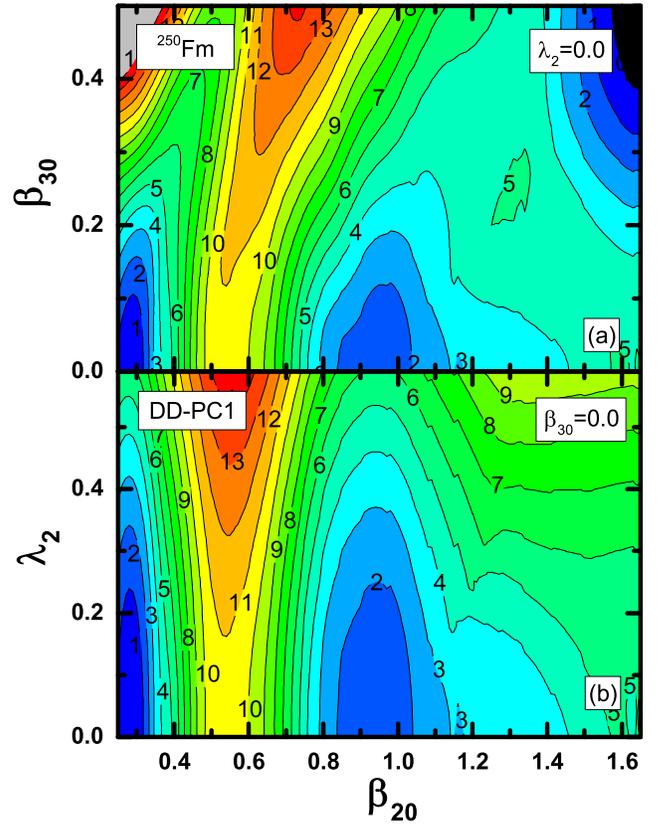}
\caption{(Color online)~\label{fig:Fm250_FTV}%
Effective collective potential $V_{\rm eff}$ of $^{250}$Fm in the $(\beta_{20},\beta_{30})$ plane for $\lambda_2 = 0$ (a) 
and the $(\beta_{20},\lambda_2)$ plane for $\beta_{30}=0$ (b).
In each panel energies are normalized with respect to the corresponding value of the equilibrium minimum, 
and contours join points on the surface with the same energy (in MeV). 
The energy surfaces are calculated with the density functionals DD-PC1~\cite{Niksic2008_PRC78-034318}, 
and the pairing interaction Eq.~(\ref{pairing}).
}
\end{figure}
%-----

Since the triaxial shape degree of freedom is suppressed, that is, it does not contribute to the 
action integral in the dynamic case when pairing fluctuations are included,  
we can analyze the SF decay of $^{250}$Fm in the restricted 3D collective space with coordinates $(\beta_{20},\beta_{30},\lambda_{2})$.
In Fig.~\ref{fig:Fm250_FTV} we display the collective potential of $^{250}$Fm in the $(\beta_{20},\beta_{30})$  plane for $\lambda_{2}=0$ (a),
and in the $(\beta_{20},\lambda_2)$ plane for $\beta_{30}=0$ (b). 
The mean-field equilibrium (ground) state is predicted at moderate
quadrupole deformation $\beta_{20}\approx 0.3$, and the isomeric minimum at 
$\beta_{20}\approx 0.95$. The nucleus remains reflection symmetric
through the entire region of quadrupole deformations $\beta_{20} \le 1.4$.
As in the previous cases, at each deformation the potential energy rises steeply with  
increasing $\lambda_{2}$, as shown in the lower panel of Fig.~\ref{fig:Fm250_FTV}.

%----
\begin{figure}
 \includegraphics[width=0.48\textwidth]{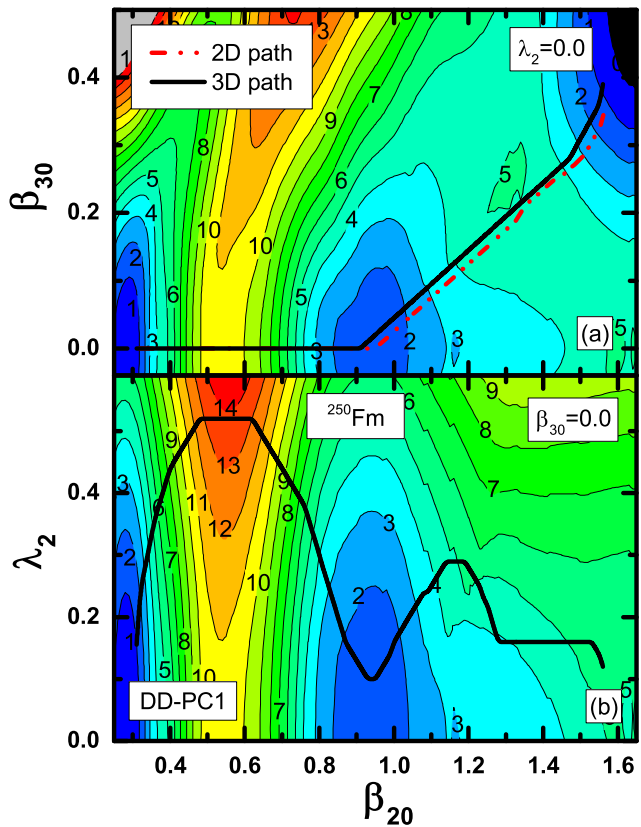}
\caption{(Color online)~\label{fig:Fm250_FT_Path}%
Projections of the 3D dynamic fission path (solid curves) 
of $^{250}$Fm on the $(\beta_{20},\beta_{30})$ plane for $\lambda_2 = 0$ (a),  
and the $(\beta_{20},\lambda_2)$ plane for $\beta_{30}=0$ (b).
The dash-dot (red) curve denotes the 2D path computed 
without the inclusion of dynamic pairing correlations.
}
\end{figure}
%----

The dynamic path computed in the restricted 2D collective space $(\beta_{20},\beta_{30})$ (dash-dot red curve), 
and the projections of the path determined in the 3D collective space 
($\beta_{20},\beta_{30},\lambda_{2}$) (solid curves), are shown in Fig.~\ref{fig:Fm250_FT_Path}.
In the $(\beta_{20},\beta_{30})$ plane the 2D and 3D paths are almost indistinguishable.
From the mean-field equilibrium state to the fission isomer the reflection-asymmetric shape degree of freedom does not contribute, 
whereas it plays a crucial role along the dynamic path connecting the isomeric state and the outer turning point. 
The effect of pairing fluctuations on the asymmetric fission of $^{250}$Fm is illustrated in the lower panel of Fig.~\ref{fig:Fm250_FT_Path}.
The dynamic pairing correlations are markedly amplified along the fission path when the nucleus traverses the inner and outer barriers, 
and this has a significant effect on the effective inertia. In fact, even though the 2D and 3D paths are almost identical in the plane 
of axially symmetric shape degrees of freedom $(\beta_{20},\beta_{30})$, because of  
pairing fluctuations in the 3D case the fission action integral is reduced by as much as 
$\sim 10$ units (see Table~\ref{tab:act}).

%--------
\begin{table}
\caption{\label{tab:act} %
Values for the action integral and SF half-lives of $^{264}$Fm and $^{250}$Fm that correspond to the 
fission paths displayed in Figs.~\ref{fig:Fm264_Path} and \ref{fig:Fm250_FT_Path}. \\
}
\begin{ruledtabular}
%\begin{tabular*}{170mm}{@{\extracolsep{\fill}}ccc}
\begin{tabular}{llcr}
%\hline \hline
Nucleus         & Path     &  S(L)        & $\log_{10}(T_{1/2}/{\rm yr})$ \\ \hline 
\\
$^{264}$Fm   & 2D        &  $19.58$  & $-11.03$ \\
		      & 3D        &  $14.15$  & $-15.75$ \\ 
$^{250}$Fm   & 2D        &  $32.09$  & $-0.16$ \\
		      & 3D        &  $22.33$  & $-8.64$ \\ 
%\hline \hline
\end{tabular}
\end{ruledtabular}
\end{table}
%--------

The calculated action integrals and resulting fission half-lives for $^{264}$Fm and $^{250}$Fm are listed in Tab.~\ref{tab:act}.
2D denotes the values obtained in two-dimensional collective spaces without taking dynamic pairing fluctuations into account,
whereas 3D labels the values calculated in three-dimensional spaces that include the pairing degree of freedom 
as collective coordinate. For the symmetric spontaneous fission of $^{264}$Fm 
the dynamic paths are determined in the $(\beta_{20},\beta_{22})$ and  $(\beta_{20},\beta_{22},\lambda_{2})$ collective spaces.
The path traverses a single fission barrier and the axial symmetry of the fissioning system 
is fully restored by the inclusion of dynamic pairing correlations. 
As a result, in the 3D space the action integral is reduced by $\sim 5$ units, and the predicted half-life 
is almost five orders of magnitude shorter than in the 2D case that neglects dynamic pairing fluctuations.
For the asymmetric fissioning nucleus $^{250}$Fm we have also shown that, although triaxial effects are important in the 
static case or in the dynamic case without pairing fluctuations, the triaxial shape degree of freedom does not play a role when 
pairing is included as collective variable, just as in the case of $^{264}$Fm.
Therefore, the SF action integral and fission half-life can be determined in the 3D collective space $(\beta_{20},\beta_{30},\lambda_{2})$.
Two barriers are traversed by the fission path and, as the result of significantly enhanced dynamic pairing correlations, 
the action integral is reduced by $\sim 10$ units as 
compared to the value computed along the 2D path in the $(\beta_{20},\beta_{30})$ collective space.
The corresponding half-life is almost nine orders of magnitude shorter than the value predicted in the 
2D space $(\beta_{20},\beta_{30})$.

%%%%%%%%%%%%%%%%%%%%%%%%%%%%%%%%%%%%%%%%%%%%%%%%%%
\section{\label{sec:summary}Summary}
%%%%%%%%%%%%%%%%%%%%%%%%%%%%%%%%%%%%%%%%%%%%%%%%%%

The dynamics of spontaneous fission of $^{264}$Fm and $^{250}$Fm 
have been investigated in a theoretical framework based on relativistic energy density functionals and,  
in addition to shape deformation degrees of freedom, pairing correlations have been included as 
collective coordinates.
Effective potentials and nonperturbative ATDHFB cranking collective inertia tensors have been 
calculated using the multidimensionally-constrained relativistic mean-field (MDC-RMF) model 
based on the energy density functional DD-PC1, and pairing correlations taken into account in 
the BCS approximation with a separable pairing force of finite range. The effect of coupling between 
shape and pairing degrees of freedom on 
dynamic (least-action) fission paths, as well as the corresponding SF half-lives has been analyzed. 

$^{264}$Fm undergoes symmetric fission into two $^{132}$Sn nuclei.
Hence, this process can be described in the 3D collective space $(\beta_{20},\beta_{22},\lambda_{2})$,
where $\lambda_{2}$ is the Lagrange multiplier related to pairing fluctuations via the particle-number dispersion 
operator. The dynamic path that connects the mean-field 
ground state and the isomeric state of $^{250}$Fm is also studied in this 3D collective space.
For both nuclei triaxial deformations reduce the height of the static inner barrier by 
$2 - 3$ MeV. However, axial symmetry of the fissioning system is fully restored along the dynamic paths 
when pairing is included as collective coordinate and, simultaneously,
pairing correlations are significantly enhanced.
The description of asymmetric spontaneous fission of $^{250}$Fm 
necessitates the inclusion of the octupole (reflection-asymmetric) 
degree of freedom $\beta_{30}$ and, in principle, calculations should be carried out in the full 
4D space spanned by the collective coordinates $(\beta_{20},\beta_{22},\beta_{30},\lambda_{2})$.
However, since the triaxial degree of freedom does not play a role in the dynamic case that includes 
pairing fluctuations, the fission action integral could be computed along the dynamic path in the symmetry-restricted 
3D collective space $(\beta_{20},\beta_{30},\lambda_{2})$.
The octupole deformation degree of freedom becomes crucial beyond the isomeric state, 
and pairing correlations display a pronounced increase when the path traverses the inner and outer barriers.
Consistent with the findings of Ref.~\cite{Sadhukhan2014_PRC90-061304}, it has been shown that the 
inclusion of pairing correlations in the space of collective coordinates, that is, the dynamical coupling between shape and 
pairing degrees of freedom, reduces the fission action integral by several units (more than five in the case 
of $^{264}$Fm, and almost ten for $^{250}$Fm) and, therefore, has a dramatic effect on the 
calculated SF half-lives. 

%%%%%%%%%%%%%%%%%%%%%%%%%%%%%%%%%%%%%%%%%%
\bigskip
%---------------------------------------------------------
\acknowledgements
This work has been supported in part by the NEWFELPRO project
of the Ministry of Science, Croatia, co-financed through the Marie
Curie FP7-PEOPLE-2011-COFUND program, 
the Croatian Science Foundation -- project ``Structure and Dynamics
of Exotic Femtosystems" (IP-2014-09-9159), the National Key Basic
Research Program of China (Grant No. 2013CB834400), the
National Natural Science Foundation of China (Grants No. 11120101005,
No. 11120101005, No. 11275248 and No. 11525524) and the Knowledge
Innovation Project of the Chinese Academy of Sciences (Grant No. KJCX2-EW-N01).
Calculations have been performed in part at  
the High-performance Computing Cluster of SKLTP/ITP-CAS.

\end{document}